\documentclass[aps,prd,twocolumn,superscriptaddress,nofootinbib,floatfix]{revtex4-2}

\usepackage{amsmath,amssymb,amsfonts}
\usepackage{bm}
\usepackage{graphicx}
\usepackage{booktabs}
\usepackage[colorlinks=true,linkcolor=blue,citecolor=blue,urlcolor=blue]{hyperref}

\newcommand{\dd}{\mathrm{d}}
\newcommand{\Lm}{\mathcal{L}_m}
\newcommand{\Pmean}{\mathcal{P}}
\newcommand{\Th}{\Theta}
\newcommand{\half}{\tfrac{1}{2}}
\newcommand{\Dden}{D}

\begin{document}

\title{Traversable wormholes in $f(T,\tau)$ gravity: \\ a complete classification of the non-exotic sector}

\author{Ayan Banerjee}
\email{ayanbanerjeemath@gmail.com}
\affiliation{Astrophysics and Cosmology Research Unit, School of Mathematics, Statistics and Computer Science, University of KwaZulu--Natal, Private Bag X54001, Durban 4000, South Africa}

\author{Takol Tangphati}
\email{takoltang@gmail.com}
\affiliation{School of Science, Walailak University, Thasala, Nakhon Si Thammarat 80160, Thailand}
\affiliation{Research Center for Theoretical Simulation and Applied Research in Bioscience and Sensing, Walailak University, Thasala, Nakhon Si Thammarat 80160, Thailand}

\author{Safiqul Islam}
\thanks{Corresponding author}
\email{sislam@kfu.edu.sa}
\affiliation{Department of Mathematics and Statistics, College of Science, King Faisal University, P.O. Box 400, Al Ahsa 31982, Saudi Arabia}

\author{Safyan Mukhtar}
\email{smahmad@kfu.edu.sa}
\affiliation{Department of Mathematics and Statistics, College of Science, King Faisal University, P.O. Box 400, Al Ahsa 31982, Saudi Arabia}

\date{\today}

\begin{abstract}
We study static and spherically symmetric traversable wormholes in
$f(T,\tau)$ gravity, where the torsion scalar $T$ is coupled to the trace
$\tau$ of the matter energy--momentum tensor. We consider the linear model
$f(T,\tau)=T+\beta\tau$ with an anisotropic fluid and the mean-pressure matter
Lagrangian $\Lm=\Pmean=(p_r+2p_t)/3$. The field equations are obtained for the
Morris--Thorne geometry without fixing the redshift or shape function at the
outset. For a constant redshift function, the energy-condition problem takes
a simple form. On the branch $\beta>8\pi$ and for $b(r)>0$, the energy density
together with the null, weak, and strong energy conditions is satisfied
throughout the spacetime if and only if $r b(r)$ is non-increasing. The same
condition also implies asymptotic flatness, $b(r)<r$ outside the throat, and
$b'(r_0)\leq -1$. The allowed geometries can therefore be written as
$b(r)=r_0^2 h(r)/r$, where $h(r_0)=1$ and $h(r)$ is positive and
non-increasing. For the representative family
$b(r)=r_0(r_0/r)^n$, the null, weak, and strong energy conditions hold for
$n\geq1$, while the dominant energy condition requires
$n\geq3(\beta-2\pi)/(\beta-6\pi)$. We also separate the physical matter from
the effective source and show how the trace coupling allows the physical
matter to remain non-exotic although the effective source violates the null
energy condition. Finally, we examine the marginal case $b(r)=r_0^2/r$ with a
non-constant redshift function. A decreasing redshift function can improve
the tangential null energy condition at the throat, but this improvement
cannot be maintained throughout an asymptotically flat exterior. These
results show that the matter--torsion coupling can support a broad class of
traversable wormholes without requiring exotic physical matter.
\end{abstract}

\maketitle

\section{Introduction}
\label{sec:intro}

Wormholes are non-trivial spacetime geometries that connect two distant
regions through a throat. Their early history goes back to the
Einstein--Rosen construction~\cite{Einstein:1935particle}, while the modern
study of traversable wormholes was established by Morris and
Thorne~\cite{Morris:1988wormholesa,Morris:1988wormholesc}. Several simple and
thin-shell constructions were subsequently developed
\cite{Visser:1989traversable,Visser:1989traversablea,Dias:2010thinshell}, and
standard reviews can be found in Refs.~\cite{Visser:1995lorentzian,Lobo:2017wormhole}.

The main difficulty in general relativity is that the flare-out condition at
the throat requires a violation of the null energy condition (NEC). The
supporting matter is therefore exotic in the usual general-relativistic
interpretation. Considerable effort has been devoted to reducing the amount
of exotic matter or confining it to a limited region
\cite{Visser:2003traversable,Nandi:2004volume}. Wormholes supported by phantom
or Chaplygin-type matter have been studied in
Refs.~\cite{Lobo:2005phantom,Lobo:2005stability,Lobo:2006chaplygin,Sushkov:2005wormholes,Gonzalez:2009wormholes,Kuhfittig:2009single},
while the Casimir effect provides another interesting source of negative
energy~\cite{Garattini:2019casimir}. Stability and junction constructions have
also been widely investigated
\cite{Lobo:2004linearized,Lobo:2005stability,Rosa:2021junction}.

Modified gravity offers a different way to address the same problem. The
additional geometrical terms can contribute to the effective source that
supports the throat, allowing the physical matter sector to satisfy the
standard energy conditions in suitable parameter ranges. This possibility has
been explored in viable $f(R)$ gravity~\cite{Samanta:2019validation},
Einstein--Gauss--Bonnet gravity~\cite{Mehdizadeh:2015einsteingaussbonnet},
extra-dimensional scenarios~\cite{Kar:2015can}, $f(Q)$ gravity
\cite{Rastgoo:2024traversable}, Rastall gravity
\cite{Moradpour:2017traversable,Halder:2019wormhole,Mustafa:2020noncommutative,Mustafa:2020stable},
and $f(R,\Lm)$ gravity~\cite{Solanki:2023wormhole,Rastgoo:2026nonexotic}.

Matter--geometry coupling has attracted particular attention in $f(R,T)$
gravity~\cite{Harko:2011gravity}. A wide variety of wormhole models have been
constructed in this framework, including static and Morris--Thorne solutions,
charged configurations, conformally symmetric geometries, and models supported
without exotic matter
\cite{Moraes:2017modeling,Zubair:2016static,Elizalde:2018wormhole,Elizalde:2019wormholes,Godani:2019static,Sahoo:2018phantom,Moraes:2019charged,Zubair:2019exact,Mishra:2020traversable,Banerjee:2020conformally,Chanda:2021morris,LuisRosa:2022nonexotica}.
More recently, variable equations of state have also been considered
\cite{Rastgoo:2025wormholes}. These studies show that a direct coupling between
matter and geometry can substantially change the material requirements for a
traversable throat.

Teleparallel gravity provides an alternative description of gravitation in
which torsion, rather than curvature, carries the gravitational interaction.
Its extensions based on a general function $f(T)$ have been studied in a wide
range of cosmological and astrophysical settings
\cite{Ferraro:2006modified,Bengochea:2008dark,Linder:2010einsteins,Cai:2015ft,Boehmer:2011existence}.
Wormhole geometries in modified teleparallel gravity were analysed in
Ref.~\cite{Bohmer:2012wormhole}, where the role of the energy conditions was
examined explicitly.

A further extension is obtained by coupling the torsion scalar to the trace of
the matter energy--momentum tensor. The resulting $f(T,\mathcal{T})$ theory was
introduced by Harko \textit{et al.}~\cite{Harko:2014fTtau}. In the present
work we denote the trace by $\tau$ to avoid confusing it with the torsion scalar
$T$. The theory has been applied to cosmology and dynamical systems
\cite{Duchaniya:2024cosmological,Mishra:2023constraining,Mishra:2024big,Mishra:2025cosmological,Zubair:2023reconstruction,Hounmenou:2025holographic,Rezaei:2020observational},
brane configurations~\cite{Moreira:2021firstorder}, gravastars
\cite{Ghosh:2020gravastars}, and compact stars
\cite{Pace:2017perturbative,Pace:2017quarka,Salako:2020study,Gudekli:2022study,Ashraf:2026impact,Alshammari:2026charged}.
Black-hole thermodynamics in related matter-coupled teleparallel models has
also been discussed in Ref.~\cite{Ahissou:2025thermodynamics}.

Wormholes in $f(T,\tau)$ gravity have received increasing attention. The
influence of GUP-corrected Casimir energy on zero-tidal-force wormholes was
studied in Ref.~\cite{Rizwan:2024influence}. Embedding class-I solutions based
on the Karmarkar condition were considered in
Ref.~\cite{Paramanik:2025embedding}, and a systematic analysis of wormholes in
the linear theory was recently presented in Ref.~\cite{Parsaei:2025wormholes}.
In that study, particular matter relations, including linear and variable
equations of state, were imposed to obtain explicit wormhole geometries. Our
approach reverses this logic: no equation of state is assumed, and the energy
conditions themselves are used to classify the admissible geometry. This leads
to a necessary-and-sufficient condition for the constant-redshift sector, the
general class $b(r)=r_0^2h(r)/r$, and a global obstruction for the marginal
member when the redshift function is allowed to vary. These results make
$f(T,\tau)$ gravity a natural setting in which to ask whether the matter
threading a wormhole can satisfy the standard energy conditions.

The purpose of the present work is to approach this question without first
choosing an equation of state or imposing a particular shape function. We
consider the linear model
\begin{equation}
 f(T,\tau)=T+\beta\tau,
 \label{eq:linear-model}
\end{equation}
with an anisotropic matter source and the mean-pressure matter Lagrangian
$\Lm=(p_r+2p_t)/3$. We first derive the field equations for a general
Morris--Thorne geometry. In the constant-redshift sector, the energy
conditions themselves select the allowed shape functions. We show that for
$\beta>8\pi$ the simple condition $(rb)'\leq0$ determines a complete class of
asymptotically flat wormholes supported by non-exotic physical matter. No
equation of state or traceless-matter condition is required. We then examine
the marginal member $b(r)=r_0^2/r$ and ask whether a non-constant redshift
function can improve its tangential null-energy behaviour away from the
throat.

The paper is organised as follows. In Sec.~\ref{sec:theory} we introduce the
$f(T,\tau)$ model and derive the wormhole field equations. The
constant-redshift solutions are studied in Sec.~\ref{sec:constant}. In
Sec.~\ref{sec:nonconstant} we examine the marginal shape function with a
non-constant redshift. The main results are summarised in
Sec.~\ref{sec:conclusions}. Throughout the paper we use geometrised units
$G=c=1$ and the metric signature $(-,+,+,+)$.

\section{$f(T,\tau)$ gravity and wormhole field equations}
\label{sec:theory}

\subsection{Basic formulation}
\label{subsec:formulation}

Teleparallel gravity is formulated in terms of a tetrad $e^{A}{}_{\mu}$, from
which the metric is constructed as
\begin{equation}
 g_{\mu\nu}=\eta_{AB}e^{A}{}_{\mu}e^{B}{}_{\nu},
 \qquad
 e\equiv\det(e^{A}{}_{\mu})=\sqrt{-g},
 \label{eq:tetrad-metric}
\end{equation}
with $\eta_{AB}=\mathrm{diag}(-1,1,1,1)$. In the pure-tetrad representation,
$\omega^{A}{}_{B\mu}=0$, and the torsion tensor is
\begin{equation}
 T^{\lambda}{}_{\mu\nu}
 =e_A{}^{\lambda}
 \left(\partial_\mu e^{A}{}_{\nu}-\partial_\nu e^{A}{}_{\mu}\right).
 \label{eq:torsion}
\end{equation}
The contorsion tensor and superpotential are
\begin{align}
 K^{\mu\nu}{}_{\rho}
 &=-\half\left(T^{\mu\nu}{}_{\rho}-T^{\nu\mu}{}_{\rho}
 -T_{\rho}{}^{\mu\nu}\right),
 \label{eq:contorsion}\\[1mm]
 S_{\rho}{}^{\mu\nu}
 &=\half\left(K^{\mu\nu}{}_{\rho}
 +\delta^\mu_\rho T^{\alpha\nu}{}_{\alpha}
 -\delta^\nu_\rho T^{\alpha\mu}{}_{\alpha}\right),
 \label{eq:superpotential}
\end{align}
so that the torsion scalar is
\begin{equation}
 T=S_{\rho}{}^{\mu\nu}T^{\rho}{}_{\mu\nu}.
 \label{eq:torsion-scalar-def}
\end{equation}

The action of $f(T,\tau)$ gravity is given by~\cite{Harko:2014fTtau}
\begin{equation}
 S=\frac{1}{16\pi}\int \dd^4x\,e\,f(T,\tau)
 +\int \dd^4x\,e\,\Lm.
 \label{eq:action}
\end{equation}
For the anisotropic source considered here,
\begin{equation}
 \Th^\mu{}_{\nu}=\mathrm{diag}(-\rho,p_r,p_t,p_t),
 \label{eq:EMT}
\end{equation}
with trace
\begin{equation}
 \tau=-\rho+p_r+2p_t.
 \label{eq:trace}
\end{equation}
We take the matter Lagrangian to be the mean pressure,
\begin{equation}
 \Lm=\Pmean\equiv\frac{p_r+2p_t}{3},
 \label{eq:Lm}
\end{equation}
which reduces to $\Lm=p$ in the isotropic limit. As in the standard
perfect-fluid treatment of trace-coupled models~\cite{Harko:2014fTtau}, we
adopt the closure prescription
$\partial^2\Pmean/\partial g^{\mu\nu}\partial g^{\alpha\beta}=0$. The
corresponding matter variation is then
\begin{equation}
 \Th^{(\mathrm{var})}_{\mu\nu}
 =-2\Th_{\mu\nu}+\Pmean g_{\mu\nu}.
 \label{eq:Theta-var-final}
\end{equation}
This is a prescription for the anisotropic matter Lagrangian rather than a
microscopic statement about every anisotropic fluid.

Variation of Eq.~\eqref{eq:action} with respect to the tetrad gives
\begin{widetext}
\begin{equation}
\begin{aligned}
&e_A{}^{\rho}S_{\rho}{}^{\mu\nu}(\partial_\mu T)f_{TT}
+e_A{}^{\rho}S_{\rho}{}^{\mu\nu}(\partial_\mu\tau)f_{T\tau}
+e^{-1}\partial_\mu\!\left(e e_A{}^{\rho}S_{\rho}{}^{\mu\nu}\right)f_T
+e_A{}^{\mu}T^{\lambda}{}_{\mu\kappa}S_{\lambda}{}^{\nu\kappa}f_T
\\
&\hspace{22mm}
-\frac14 e_A{}^{\nu}f
-\frac{f_\tau}{2}\left(e_A{}^{\lambda}\Th_{\lambda}{}^{\nu}
+\Pmean e_A{}^{\nu}\right)
=-4\pi e_A{}^{\lambda}\Th_{\lambda}{}^{\nu}.
\end{aligned}
\label{eq:master}
\end{equation}
\end{widetext}
For the linear model in Eq.~\eqref{eq:linear-model},
$f_T=1$, $f_\tau=\beta$, and $f_{TT}=f_{T\tau}=0$. The field equations reduce
to
\begin{equation}
 G^\mu{}_{\nu}
 =(8\pi-\beta)\Th^\mu{}_{\nu}
 -\beta\left(\Pmean+\frac{\tau}{2}\right)\delta^\mu{}_{\nu},
 \label{eq:reduced}
\end{equation}
where
\begin{equation}
 \Pmean+\frac{\tau}{2}
 =\frac{-3\rho+5p_r+10p_t}{6}.
 \label{eq:Pplustau}
\end{equation}
The general-relativistic equations are recovered continuously in the limit
$\beta\rightarrow0$. For this linear choice the geometric sector is TEGR,
while the modification enters through the trace-coupled source. The structure
is therefore analogous to, but not identical with, the linear $f(R,T)$ case
\cite{Harko:2011gravity}, because the matter variation leads to the specific
coefficients appearing in Eq.~\eqref{eq:reduced}.

The physical matter tensor is not separately conserved when $\beta\neq0$.
Taking the covariant derivative of Eq.~\eqref{eq:reduced} and using the
Bianchi identity gives
\begin{equation}
 \nabla_\mu\Th^\mu{}_{\nu}
 =\frac{\beta}{8\pi-\beta}
 \partial_\nu\left(\Pmean+\frac{\tau}{2}\right).
 \label{eq:nonconservation}
\end{equation}
Thus the trace coupling describes an exchange of energy--momentum between the
physical matter and the gravitational sector, as is typical of
matter--geometry coupled theories.

The diagonal tetrad is consistent for this linear model. For a static and
spherically symmetric spacetime, the off-diagonal equation reduces to
\begin{equation}
 f_{TT}T'+f_{T\tau}\tau'=0,
 \label{eq:offdiag}
\end{equation}
which is identically satisfied because $f_{TT}=f_{T\tau}=0$. In the
pure-tetrad representation the diagonal spherical tetrad carries the familiar
inertial torsion contribution even in the Minkowski limit, but this produces
no additional field-equation constraint for the present linear model precisely
because $f_{TT}=f_{T\tau}=0$.

\subsection{Wormhole geometry and field equations}
\label{subsec:geometry}

We consider the Morris--Thorne line element
\begin{equation}
 \dd s^2=-e^{2\Phi(r)}\dd t^2
 +\frac{\dd r^2}{1-b(r)/r}
 +r^2\left(\dd\theta^2+\sin^2\theta\,\dd\varphi^2\right),
 \label{eq:MT-metric}
\end{equation}
where $\Phi(r)$ and $b(r)$ are the redshift and shape functions,
respectively. A throat at $r=r_0$ satisfies
\begin{equation}
 b(r_0)=r_0,
 \qquad
 b'(r_0)<1,
 \qquad
 b(r)<r \quad (r>r_0),
 \label{eq:throat-conditions}
\end{equation}
and $\Phi(r)$ must remain finite to avoid a horizon. Asymptotic flatness
requires $b(r)/r\rightarrow0$ and $\Phi(r)\rightarrow\mathrm{const}$ as
$r\rightarrow\infty$.

A diagonal tetrad compatible with Eq.~\eqref{eq:MT-metric} is
\begin{equation}
 e^{A}{}_{\mu}
 =\mathrm{diag}\left(
 e^{\Phi},\frac{1}{\sqrt{1-b/r}},r,r\sin\theta\right),
 \label{eq:tetrad}
\end{equation}
which gives
\begin{equation}
 T(r)
 =-\frac{2}{r}\left(1-\frac{b}{r}\right)
 \left(2\Phi'+\frac1r\right).
 \label{eq:torsion-scalar}
\end{equation}

Substituting the metric into Eq.~\eqref{eq:reduced} and solving for the
anisotropic source yields
\begin{widetext}
\begin{align}
\rho &= \frac{1}{\Dden r^2}\Big[
-10\beta r(r-b)\Phi'^2
+5\beta\Phi'\{3b+r(b'-4)\}
-16(\beta-3\pi)b'
-10\beta r(r-b)\Phi''\Big],
\label{eq:rho}\\[2mm]
p_r &= \frac{1}{\Dden r^3}\Big[
10\beta r^3\Phi'^2-8\beta r b'
+r^2\Phi'(96\pi-28\beta-5\beta b')
+10\beta r^3\Phi''
\nonumber\\
&\hspace{22mm}
-b\{48\pi-24\beta+(96\pi-33\beta)r\Phi'
+10\beta r^2\Phi'^2+10\beta r^2\Phi''\}\Big],
\label{eq:pr}\\[2mm]
p_t &= \frac{1}{\Dden r^3}\Big[
b\{-12(\beta-2\pi)-3(\beta+8\pi)r\Phi'
+2(7\beta-24\pi)r^2(\Phi'^2+\Phi'')\}
\nonumber\\
&\hspace{22mm}
+r\{-2(7\beta-24\pi)r^2(\Phi'^2+\Phi'')
+4(\beta-6\pi)b'
+r\Phi'[48\pi-4\beta+(7\beta-24\pi)b']\}\Big],
\label{eq:pt}
\end{align}
\end{widetext}
where
\begin{equation}
 \Dden=24(\beta-8\pi)(\beta-2\pi).
 \label{eq:Ddef}
\end{equation}
These expressions determine the matter source for any Morris--Thorne geometry
in the linear theory, provided $\beta\neq2\pi,8\pi$.

The combinations entering the NEC are much simpler:
\begin{align}
\rho+p_r
&=\frac{b(1+2r\Phi')-r(2r\Phi'+b')}
{(\beta-8\pi)r^3},
\label{eq:NEC-radial}\\[2mm]
\rho+p_t
&=\frac{1}{2(\beta-8\pi)r^3}
\Big[b\{-1+r\Phi'+2r^2\Phi'^2+2r^2\Phi''\}
\nonumber\\
&\hspace{8mm}
-r\{2r^2\Phi'^2-r\Phi'(b'-2)+b'+2r^2\Phi''\}\Big].
\label{eq:NEC-tangential}
\end{align}
The trace-dependent term in Eq.~\eqref{eq:reduced} is proportional to
$\delta^\mu{}_{\nu}$ and therefore cancels when two diagonal components are
subtracted. Thus
\begin{equation}
 \rho+p_r=\frac{G^t{}_t-G^r{}_r}{\beta-8\pi},
 \qquad
 \rho+p_t=\frac{G^t{}_t-G^\theta{}_\theta}{\beta-8\pi}.
 \label{eq:NEC-structural}
\end{equation}
At the throat, Eq.~\eqref{eq:NEC-radial} reduces to
\begin{equation}
 \left.(\rho+p_r)\right|_{r_0}
 =\frac{1-b'(r_0)}{(\beta-8\pi)r_0^2}.
 \label{eq:throat}
\end{equation}
The derivative $\Phi'(r_0)$ cancels exactly. Since the flare-out condition
gives $1-b'(r_0)>0$, the radial NEC can be satisfied at a genuine throat only
for
\begin{equation}
 \beta>8\pi.
 \label{eq:beta-branch}
\end{equation}
For $\beta<8\pi$, including the GR limit $\beta\rightarrow0$, the radial NEC
is necessarily violated at the throat. The values $\beta=2\pi$ and
$\beta=8\pi$ correspond to degeneracies of the algebraic reconstruction and
are not included in the following analysis.

\section{Wormhole solutions with a constant redshift function}
\label{sec:constant}

We now consider the zero-tidal-force case. A finite constant redshift can be
absorbed into a redefinition of the time coordinate, so we set $\Phi=0$.
Equations~\eqref{eq:rho}--\eqref{eq:pt} become
\begin{align}
\rho
&=-\frac{2(\beta-3\pi)b'}
{3(\beta-8\pi)(\beta-2\pi)r^2},
\label{eq:r1-rho}\\[1mm]
p_r
&=\frac{3(\beta-2\pi)b-\beta r b'}
{3(\beta-8\pi)(\beta-2\pi)r^3},
\label{eq:r1-pr}\\[1mm]
p_t
&=\frac{-3(\beta-2\pi)b+(\beta-6\pi)r b'}
{6(\beta-8\pi)(\beta-2\pi)r^3}.
\label{eq:r1-pt}
\end{align}
The trace is
\begin{equation}
 \tau=\frac{2(\beta-6\pi)b'}
 {3(\beta-8\pi)(\beta-2\pi)r^2}.
 \label{eq:r1-tau}
\end{equation}
Thus the trace is generally non-zero and the coupling that defines the theory
remains active on the solutions considered below.

For an anisotropic fluid, the NEC requires $\rho+p_r\geq0$ and
$\rho+p_t\geq0$. The weak energy condition (WEC) adds $\rho\geq0$, the strong
energy condition (SEC) adds $\rho+p_r+2p_t\geq0$, and the dominant energy
condition (DEC) requires $\rho\geq|p_r|$ and $\rho\geq|p_t|$.

\subsection{Energy conditions and admissible shape functions}
\label{subsec:constant-general}

For constant redshift,
\begin{equation}
 \rho+p_r=\frac{b-rb'}{(\beta-8\pi)r^3},
 \qquad
 \rho+p_t=-\frac{b+rb'}{2(\beta-8\pi)r^3}.
 \label{eq:r1-nec}
\end{equation}
We restrict attention to the physically relevant branch $\beta>8\pi$. Then
\begin{equation}
 \rho\geq0 \Longleftrightarrow b'\leq0,
 \qquad
 \rho+p_r\geq0 \Longleftrightarrow b'\leq\frac{b}{r},
 \label{eq:r1-first-two}
\end{equation}
whereas the tangential NEC gives
\begin{equation}
 b+rb'\leq0
 \qquad\Longleftrightarrow\qquad
 (rb)'\leq0.
 \label{eq:r1-conditions}
\end{equation}
For a positive shape function, $b(r)>0$, this is the strongest condition. It
implies $b'\leq-b/r<0$, so $\rho\geq0$, while
$b-rb'\geq2b>0$ guarantees the radial NEC. The SEC is also non-negative under
the same restriction.

Hence, for $\beta>8\pi$ and constant redshift, a positive Morris--Thorne shape
function satisfies the energy density, NEC, WEC, and SEC throughout the
exterior if and only if $r b(r)$ is non-increasing.

This condition also fixes the asymptotic behaviour. Integrating
$(rb)'\leq0$ from the throat gives
\begin{equation}
 r b(r)\leq r_0^2,
\end{equation}
so that
\begin{equation}
 b(r)\leq\frac{r_0^2}{r},
 \qquad
 \frac{b(r)}{r}\leq\left(\frac{r_0}{r}\right)^2.
 \label{eq:r1-corollary}
\end{equation}
Therefore $b(r)/r\rightarrow0$ as $r\rightarrow\infty$, $b(r)<r$ for every
$r>r_0$, and
\begin{equation}
 b'(r_0)\leq-1,
\end{equation}
which is stronger than the usual flare-out requirement. In this sector,
asymptotic flatness is not an additional assumption: it follows from the
requirement that the physical matter be non-exotic. A further consequence is
that $b(r)\rightarrow0$ at spatial infinity. Hence the ADM mass of every
geometry in this admissible constant-redshift class vanishes,
\begin{equation}
 M_{\rm ADM}=\frac12\lim_{r\to\infty}b(r)=0.
 \label{eq:ADM-zero}
\end{equation}
The zero asymptotic mass will be consistent with the effective-source picture
discussed below, where the positive physical matter contribution is modified
by the trace coupling.

The complete class can be written as
\begin{equation}
 b(r)=\frac{r_0^2}{r}h(r),
 \label{eq:r1-h}
\end{equation}
with
\begin{equation}
 h(r_0)=1,
 \qquad h(r)>0,
 \qquad h'(r)\leq0.
 \label{eq:r1-h-bijection}
\end{equation}
Thus the energy conditions determine a family of admissible geometries rather
than merely constraining a particular ansatz.

\subsection{Power-law solutions and physical properties}
\label{subsec:powerlaw}

A convenient representative is
\begin{equation}
 b(r)=r_0\left(\frac{r_0}{r}\right)^n,
 \qquad n>0.
 \label{eq:r1-shape}
\end{equation}
For this family,
\begin{equation}
 (rb)'=-(n-1)\frac{r_0^{n+1}}{r^n},
\end{equation}
so Eq.~\eqref{eq:r1-conditions} requires $n\geq1$. The throat condition is
satisfied with $b'(r_0)=-n$, and
$b(r)/r=(r_0/r)^{n+1}\rightarrow0$ at large radius.

The density and trace are
\begin{align}
 \rho&=\frac{2n(\beta-3\pi)b}
 {3(\beta-8\pi)(\beta-2\pi)r^3},
 \label{eq:r1-fam-rho}\\[1mm]
 \tau&=-\frac{2n(\beta-6\pi)b}
 {3(\beta-8\pi)(\beta-2\pi)r^3}.
 \label{eq:r1-fam-tau}
\end{align}
while
\begin{equation}
 \rho+p_r=\frac{(n+1)b}{(\beta-8\pi)r^3},
 \qquad
 \rho+p_t=\frac{(n-1)b}{2(\beta-8\pi)r^3}.
 \label{eq:r1-fam-nec}
\end{equation}
For $\beta>8\pi$ and $n\geq1$, the density is positive and both null
combinations are non-negative for all $r\geq r_0$. The WEC and SEC are
therefore satisfied throughout the spacetime. The limiting member $n=1$ gives
$b(r)=r_0^2/r$ and saturates the tangential NEC,
$\rho+p_t\equiv0$. The matter--geometry exchange is non-zero on this family;
Eq.~\eqref{eq:nonconservation} reduces to
\begin{equation}
 \nabla_\mu\Th^\mu{}_{r}
 =-\frac{\beta n(n+3)r_0^{n+1}}
 {3(\beta-8\pi)(\beta-2\pi)r^{n+4}},
 \label{eq:powerlaw-nonconservation}
\end{equation}
which vanishes in the GR limit but not on the branch considered here.

\begin{figure*}[t]
\centering
\includegraphics[width=\textwidth]{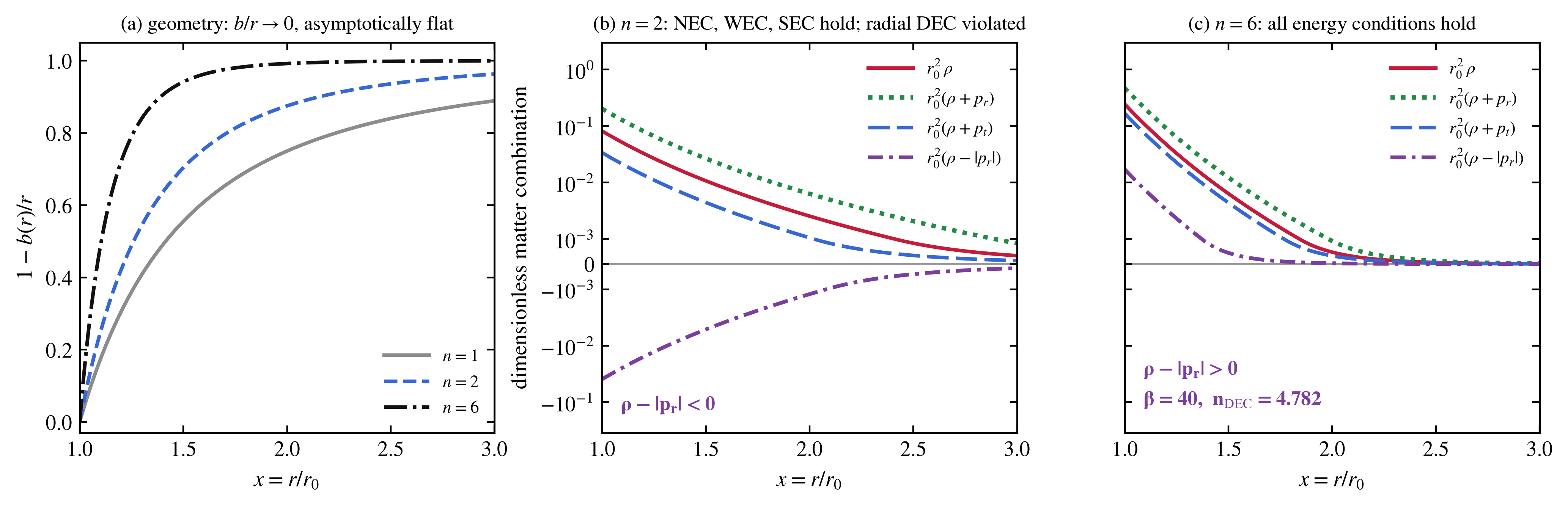}
\caption{Geometry and matter variables for
$b(r)=r_0(r_0/r)^n$ at $\beta=40$. Panel (a) shows $1-b/r$ for representative
values of $n$ and confirms the asymptotically flat behaviour. Panel (b)
corresponds to $n=2$, for which the NEC, WEC, and SEC are satisfied but the
radial DEC is violated. Panel (c) shows $n=6$, where all four standard energy
conditions are satisfied.}
\label{fig:r1-profiles}
\end{figure*}

Figure~\ref{fig:r1-profiles} illustrates the hierarchy of the energy
conditions for representative members of the power-law family. Panel (a)
confirms the asymptotically flat behaviour of the geometry. For $n=2$, the
density and both null combinations remain positive, but the radial DEC is
violated. By contrast, $n=6$ lies above the analytical threshold
$n_{\rm DEC}\simeq4.782$ at $\beta=40$ and satisfies all four standard
energy conditions. The numerical profiles are therefore consistent with
the analytical bounds derived below.

The DEC gives one additional restriction. For $\beta>8\pi$, the radial
pressure is positive and the tangential pressure is negative. The tangential
DEC therefore reduces to $\rho+p_t\geq0$ and introduces no new bound. The
radial condition $\rho-p_r\geq0$ gives
\begin{equation}
 n\geq n_{\rm DEC}(\beta)
 =\frac{3(\beta-2\pi)}{\beta-6\pi}.
 \label{eq:r1-dec}
\end{equation}
The threshold decreases from $9$ as $\beta\rightarrow8\pi^+$ and approaches
$3$ as $\beta\rightarrow\infty$. For $\beta=40$,
$n_{\rm DEC}\simeq4.782$. Equivalently,
\begin{equation}
 \omega\equiv\frac{p_r}{\rho}
 =\frac{\beta(3+n)-6\pi}{2n(\beta-3\pi)},
 \label{eq:r1-omega}
\end{equation}
and the radial DEC is simply $\omega\leq1$. Here $\omega$ is used only to
characterise the reconstructed static source; no sound-speed or stability
interpretation is attached to it.

\begin{figure*}[!t]
\centering
\includegraphics[width=\textwidth]{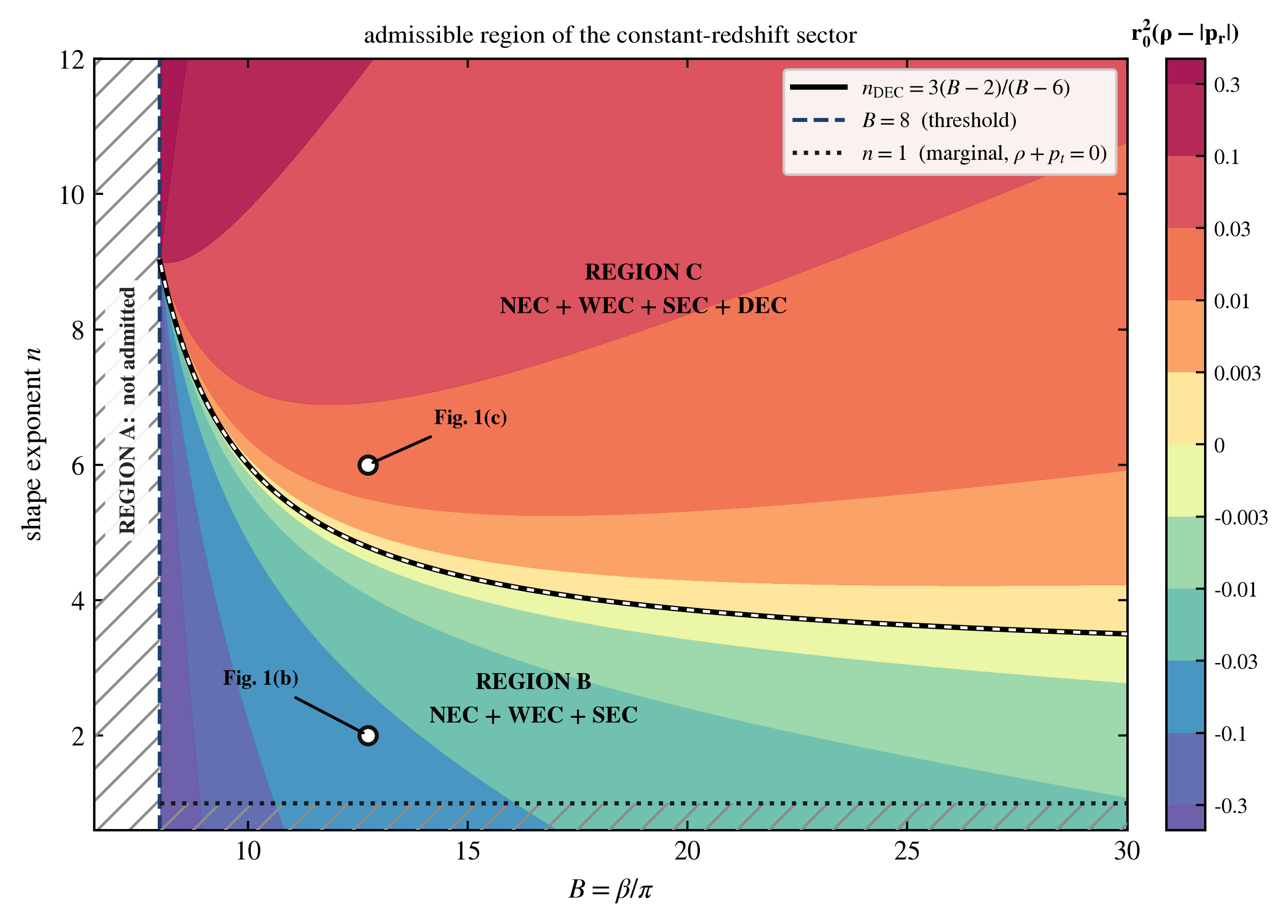}
\caption{Allowed region in the $(B,n)$ plane, where $B=\beta/\pi$. The
colour field is the dimensionless radial-DEC margin
$r_0^2(\rho-|p_r|)$ evaluated at the throat $r=r_0$. The hatched region
$B\leq8$ is excluded by the radial NEC, while the hatched strip $n<1$ for
$B>8$ lies outside the class satisfying the NEC, WEC, and SEC. For $B>8$ and
$n\geq1$, the NEC, WEC, and SEC are satisfied, and above the curve
$n_{\rm DEC}=3(B-2)/(B-6)$ the DEC is satisfied as well.}
\label{fig:r1-map}
\end{figure*}

Figure~\ref{fig:r1-map} summarises the analytical restrictions in the
$(B,n)$ plane. The line $B=8$ separates the branch where the radial NEC
necessarily fails at the throat from the non-exotic branch studied here.
For $B>8$, the region $n\geq1$ satisfies the NEC, WEC, and SEC, whereas
the stronger DEC requirement selects the region above
$n_{\rm DEC}=3(B-2)/(B-6)$. The parameter map therefore provides a direct
visual check of the bounds derived from the field equations.

It is useful to distinguish the physical matter from the effective source
seen by the Einstein tensor. Writing
$G^\mu{}_{\nu}=8\pi\Th^\mu{}_{\nu,\mathrm{eff}}$, one obtains
\begin{equation}
 \Th^\mu{}_{\nu,\mathrm{eff}}
 =\left(1-\frac{\beta}{8\pi}\right)\Th^\mu{}_{\nu}
 -\frac{\beta}{8\pi}
 \left(\Pmean+\frac{\tau}{2}\right)\delta^\mu{}_{\nu}.
 \label{eq:r1-eff-def}
\end{equation}
For either null direction,
\begin{equation}
 (\rho+p_i)_{\rm eff}
 =\left(1-\frac{\beta}{8\pi}\right)(\rho+p_i),
 \qquad i=r,t.
 \label{eq:r1-eff-null}
\end{equation}
Since $\beta>8\pi$, the physical and effective null combinations have
opposite signs. For the power-law family,
\begin{equation}
 \rho_{\rm eff}=-\frac{n b}{8\pi r^3},
 \qquad
 p_{r,\rm eff}=-\frac{b}{8\pi r^3},
 \qquad
 p_{t,\rm eff}=\frac{(n+1)b}{16\pi r^3},
 \label{eq:r1-eff-components}
\end{equation}
and
\begin{align}
 (\rho+p_r)_{\rm eff}
 &=-\frac{(n+1)b}{8\pi r^3}<0,
 \label{eq:r1-eff-nec-r}\\[1mm]
 (\rho+p_t)_{\rm eff}
 &=-\frac{(n-1)b}{16\pi r^3}\leq0.
 \label{eq:r1-eff-nec-t}
\end{align}
The effective source retains the exotic behaviour required by the wormhole
geometry, while the physical matter can satisfy the standard energy
conditions. The trace coupling controls the mapping between these two
descriptions.

The same sign reversal also has a broader physical implication for the
strong-coupling branch. For pressureless matter, $p_r=p_t=0$, one has
$\Pmean=0$ and $\tau=-\rho$, so Eq.~\eqref{eq:r1-eff-def} gives
\begin{equation}
 \rho_{\rm eff}
 =\left(1-\frac{3\beta}{16\pi}\right)\rho.
 \label{eq:dust-effective}
\end{equation}
Thus positive-density dust has a negative effective density throughout the
wormhole branch $\beta>8\pi$. The non-exotic solutions should therefore be
interpreted with this qualification: the coupling does not remove the
exoticity required by the geometry, but relocates it to the effective
gravity sector. This opposite-sign effective response also indicates that the
weak-field viability of the $\beta>8\pi$ branch is a separate question and
cannot be inferred from the present wormhole analysis alone.

\begin{figure*}[!t]
\centering
\includegraphics[width=\textwidth]{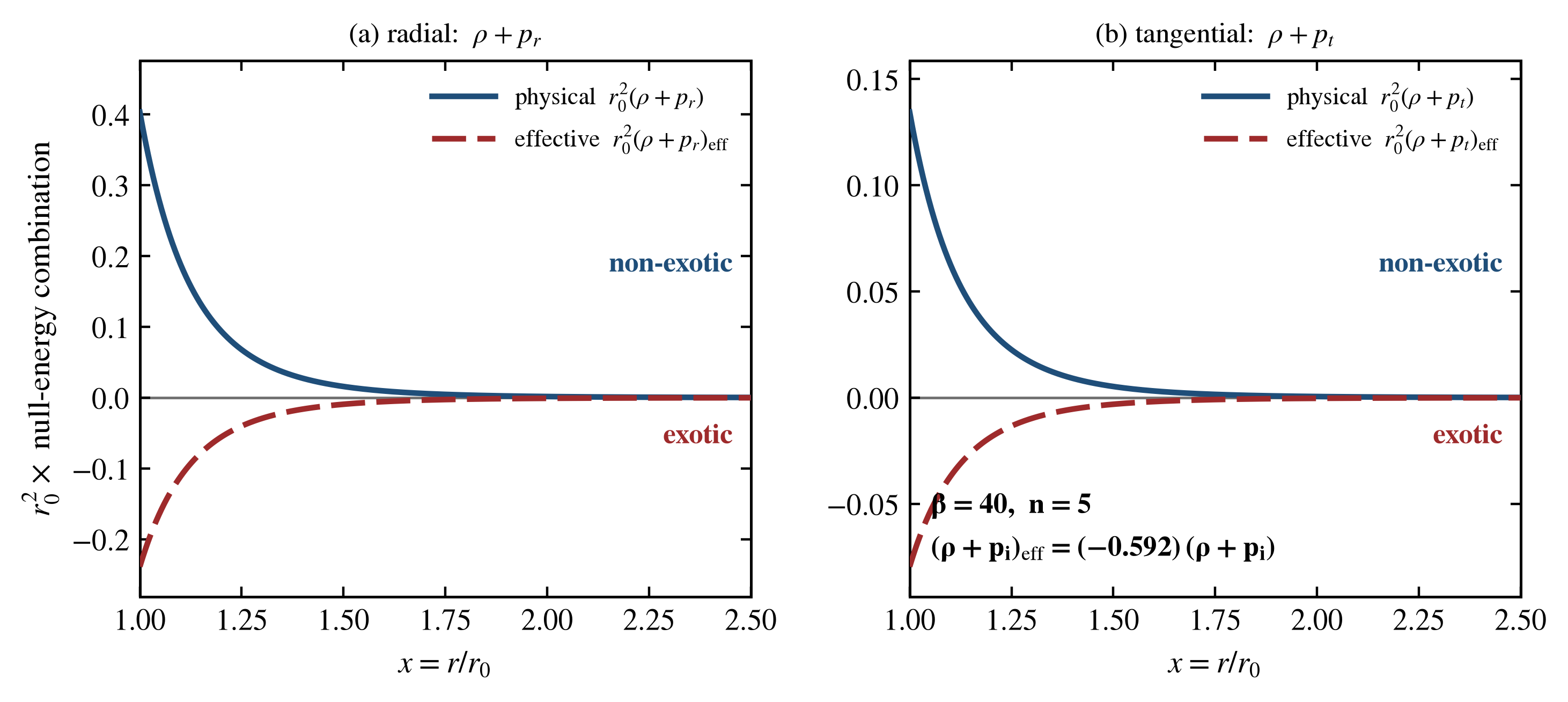}
\caption{Comparison between the physical and effective null-energy
combinations for $\beta=40$ and $n=5$. The physical matter satisfies both null
conditions, whereas the effective source violates them. The sign reversal
follows directly from Eq.~\eqref{eq:r1-eff-null}.}
\label{fig:r1-effective}
\end{figure*}

The behaviour shown in Fig.~\ref{fig:r1-effective} makes the supporting
mechanism transparent for the explicit choice $\beta=40$ and $n=5$. Both
physical null-energy combinations remain positive, while their effective
counterparts are negative. Since $1-\beta/(8\pi)<0$ on the branch
$\beta>8\pi$, this sign reversal follows directly from
Eq.~\eqref{eq:r1-eff-null}. The exoticity required by the wormhole geometry
is therefore carried by the effective gravitational source rather than by the
physical anisotropic matter.

The geometry is also regular at the throat. For the family
\eqref{eq:r1-shape},
\begin{align}
 R&=-\frac{2n r_0^{n+1}}{r^{n+3}},
 \label{eq:r1-R}\\[1mm]
 R_{\mu\nu}R^{\mu\nu}
 &=\frac{3n^2+2n+3}{2}\frac{r_0^{2n+2}}{r^{2n+6}},
 \label{eq:r1-RicSq}\\[1mm]
 K&\equiv R_{\mu\nu\rho\sigma}R^{\mu\nu\rho\sigma}
 =2(n^2+2n+3)\frac{r_0^{2n+2}}{r^{2n+6}}.
 \label{eq:r1-K}
\end{align}
All three invariants remain finite at $r=r_0$ and vanish at infinity. The
proper radial distance is
\begin{equation}
 \ell(r)=\pm\int_{r_0}^{r}
 \frac{\dd u}{\sqrt{1-(r_0/u)^{n+1}}},
 \label{eq:r1-proper}
\end{equation}
which is finite at the throat. For $n=1$ it reduces to
$\ell(r)=\pm\sqrt{r^2-r_0^2}$. In the constant-redshift sector the radial
tidal term vanishes identically, while the remaining lateral tidal effect is
finite for a finite throat radius. The solutions are therefore regular,
horizon-free, and geometrically traversable.

\section{Wormholes with a non-constant redshift function}
\label{sec:nonconstant}

The member $n=1$ lies exactly at the boundary of the allowed
constant-redshift family because $\rho+p_t=0$. This makes
\begin{equation}
 b(r)=\frac{r_0^2}{r}
 \label{eq:r2-shape}
\end{equation}
a natural case in which to ask whether a varying redshift function can turn
the saturated tangential NEC into a strict inequality.

\subsection{General behaviour of the null energy condition}
\label{subsec:nogo}

Substitution of Eq.~\eqref{eq:r2-shape} into the general null-energy
expressions gives
\begin{align}
\rho+p_r
&=\frac{2\left[r_0^2-r(r^2-r_0^2)\Phi'\right]}
{(\beta-8\pi)r^4},
\label{eq:r2-necr}\\[2mm]
\rho+p_t
&=\frac{(r_0^2-r^2)(\Phi'^2+\Phi'')-r\Phi'}
{(\beta-8\pi)r^2}.
\label{eq:r2-nect}
\end{align}
At the throat,
\begin{equation}
 \left.(\rho+p_r)\right|_{r_0}
 =\frac{2}{(\beta-8\pi)r_0^2},
\end{equation}
while
\begin{equation}
 \left.(\rho+p_t)\right|_{r_0}
 =-\frac{\Phi'(r_0)}{(\beta-8\pi)r_0}.
 \label{eq:r2-throat-t}
\end{equation}
Thus, for $\beta>8\pi$, a decreasing redshift function
$\Phi'(r_0)<0$ makes the tangential NEC strictly positive at the throat. The
remaining question is whether this improvement can hold throughout the
exterior.

Define
\begin{equation}
 Y(r)=e^{\Phi(r)}>0,
 \qquad
 F(r)=\sqrt{r^2-r_0^2}\,Y'(r).
 \label{eq:r2-F}
\end{equation}
Using $Y''=(\Phi''+\Phi'^2)Y$, Eq.~\eqref{eq:r2-nect} becomes
\begin{equation}
 \rho+p_t
 =-\frac{\sqrt{r^2-r_0^2}}
 {(\beta-8\pi)r^2Y(r)}F'(r).
 \label{eq:r2-nect-F}
\end{equation}
For $\beta>8\pi$ and $r>r_0$,
\begin{equation}
 \rho+p_t\geq0
 \qquad\Longleftrightarrow\qquad
 F'(r)\leq0.
 \label{eq:r2-equiv}
\end{equation}
At the throat $F(r_0)=0$. For an asymptotically flat redshift function with
$r\Phi'(r)\rightarrow0$, one also has $F(\infty)=0$. Therefore
\begin{equation}
 \int_{r_0}^{\infty}F'(r)\,\dd r
 =F(\infty)-F(r_0)=0.
 \label{eq:r2-integral}
\end{equation}
If the tangential NEC were satisfied everywhere, then $F'(r)\leq0$ over the
whole exterior. Together with Eq.~\eqref{eq:r2-integral}, this requires
$F'(r)=0$, hence $F(r)=0$ and finally $\Phi'(r)=0$. We therefore conclude that
for the marginal shape function $b(r)=r_0^2/r$, an everywhere finite,
asymptotically flat, non-constant redshift function cannot satisfy the
tangential NEC throughout the entire exterior. A negative $\Phi'(r_0)$ can
improve the condition locally at the throat, but the gain must be lost at
some larger radius.

\subsection{An illustrative redshift model}
\label{subsec:redshift-example}

To illustrate the above result, consider
\begin{equation}
 \Phi(r)=\frac{a r_0}{r},
 \qquad a>0.
 \label{eq:r2-phi-family}
\end{equation}
This function is finite everywhere, approaches zero at infinity, and satisfies
$\Phi'(r_0)=-a/r_0<0$. The relevant matter variables are
\begin{align}
\rho
&=\frac{r_0^2\left[5a\beta r r_0+8(\beta-3\pi)r^2
-5a^2\beta(r^2-r_0^2)\right]}
{12(\beta-8\pi)(\beta-2\pi)r^6},
\label{eq:r2-fam-rho}\\[2mm]
\rho+p_r
&=\frac{2r_0\left[a(r^2-r_0^2)+r r_0\right]}
{(\beta-8\pi)r^5},
\label{eq:r2-fam-necr}\\[2mm]
\rho+p_t
&=-\frac{a r_0\left[r(r^2-2r_0^2)+a r_0(r^2-r_0^2)\right]}
{(\beta-8\pi)r^6}.
\label{eq:r2-fam-nect}
\end{align}
For $\beta>8\pi$ and $a>0$, the radial null combination is positive. The
tangential combination changes sign at the root $x=r/r_0>1$ of
\begin{equation}
 x^3+a x^2-2x-a=0.
 \label{eq:r2-root}
\end{equation}
The crossing occurs at $r/r_0\simeq1.358$, $1.313$, $1.247$, and $1.170$ for
$a=1/4$, $1/2$, $1$, and $2$, respectively. Increasing the redshift gradient
therefore improves the tangential NEC at the throat but causes it to fail
closer to the throat outside.

A second restriction comes from the energy density. At large $r$ its sign is
controlled by $(8-5a^2)\beta-24\pi$, giving the critical value
\begin{equation}
 a_{\rm crit}(\beta)
 =2\sqrt{\frac{2}{5}}\sqrt{1-\frac{3\pi}{\beta}}.
 \label{eq:r2-acrit}
\end{equation}
For $a>a_{\rm crit}$, the density eventually becomes negative. At
$\beta=40$, $a_{\rm crit}\simeq1.106$, while for large $\beta$ it approaches
$2\sqrt{2/5}\simeq1.265$.

\begin{figure*}[!t]
\centering
\includegraphics[width=\textwidth]{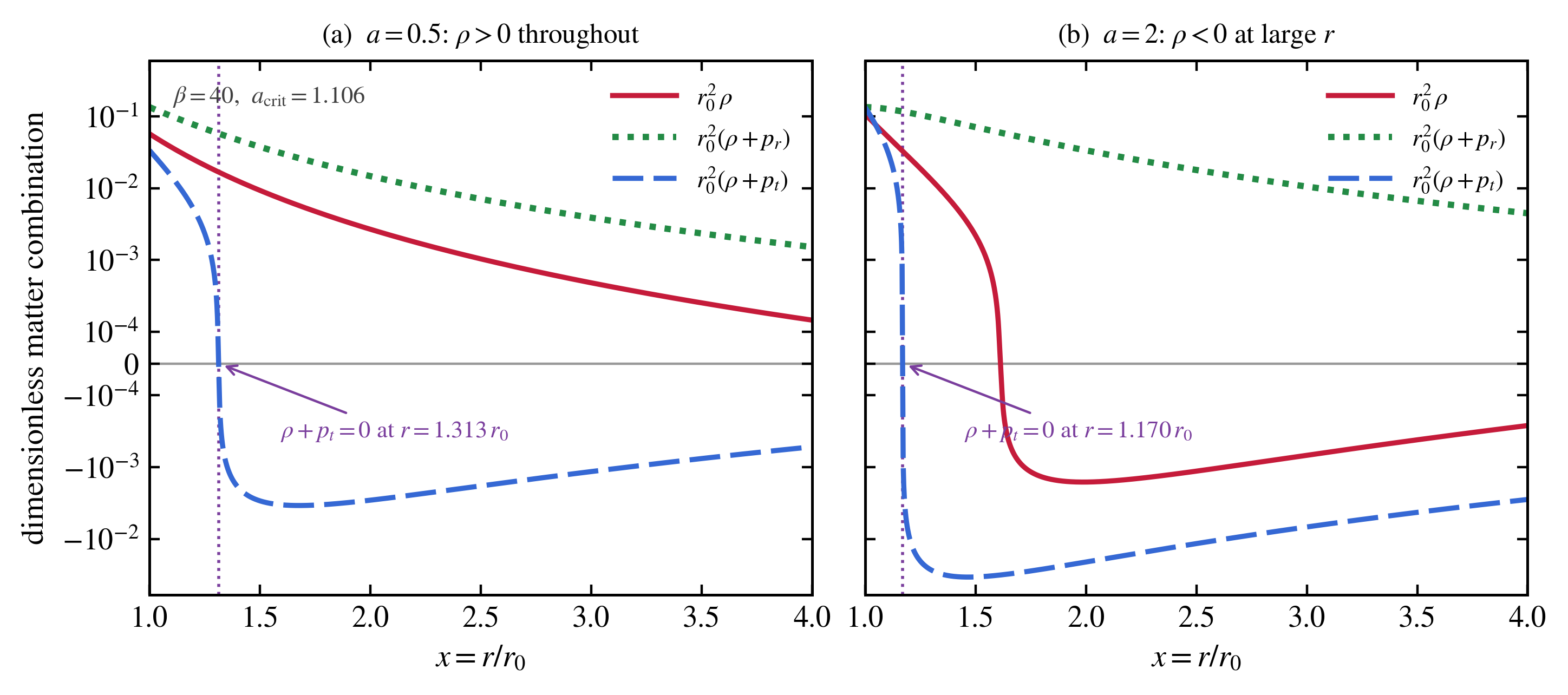}
\caption{Energy-condition behaviour for $b(r)=r_0^2/r$,
$\Phi(r)=a r_0/r$, and $\beta=40$. For $a=1/2<a_{\rm crit}$, the density and
radial NEC remain positive, whereas the tangential NEC changes sign at
$r\simeq1.313r_0$. For $a=2>a_{\rm crit}$, the tangential NEC fails closer to
the throat, at $r\simeq1.170r_0$, and the density also becomes negative at
larger radius.}
\label{fig:r2-nogo}
\end{figure*}

Figure~\ref{fig:r2-nogo} displays the two limitations of the
non-constant-redshift construction. For $a=1/2<a_{\rm crit}$, the density
and radial NEC remain positive, but the tangential NEC changes sign at
$r\simeq1.313\,r_0$, showing explicitly that the throat-level improvement
cannot persist throughout the exterior. For $a=2>a_{\rm crit}$, the
tangential NEC fails even closer to the throat and the density becomes
negative at larger radius. The second panel therefore also illustrates the
loss of the WEC when the redshift gradient is sufficiently large.

The explicit model confirms the general result. A non-constant redshift can
improve the tangential NEC locally, but it does not convert the marginal
$n=1$ geometry into an everywhere strictly non-exotic solution. For
sufficiently large redshift gradients, the WEC also fails because the energy
density becomes negative.

\section{Summary and conclusions}
\label{sec:conclusions}

We have investigated static and spherically symmetric traversable wormholes
in the linear matter-coupled teleparallel model
$f(T,\tau)=T+\beta\tau$, using an anisotropic source with the mean-pressure
matter Lagrangian $\Lm=(p_r+2p_t)/3$. The field equations were first obtained
for a general Morris--Thorne geometry, so the role of the redshift function
could be separated clearly from the conditions imposed by the throat. A key
result is that the radial null energy condition at the throat depends on the
coupling through the factor $\beta-8\pi$. Since the flare-out condition makes
the numerator positive, the branch that allows non-exotic physical matter is
$\beta>8\pi$. Importantly, this conclusion does not depend on
$\Phi'(r_0)$.

The constant-redshift sector gives the clearest result of the analysis. For
a positive shape function, the requirement that the physical matter satisfy
the density condition together with the NEC, WEC, and SEC reduces to the
single inequality $(rb)'\leq0$. This condition does more than constrain the
matter sector. It also forces the shape function to decrease sufficiently
rapidly that the wormhole becomes asymptotically flat, remains below
$b(r)=r$ outside the throat, and automatically satisfies the flare-out
condition. In this sense, the same requirement that removes the need for
exotic physical matter also selects a well-behaved wormhole geometry.

The admissible class can be represented by $b(r)=r_0^2h(r)/r$, with
$h(r_0)=1$, $h>0$, and $h'\leq0$. The power-law choice
$b(r)=r_0(r_0/r)^n$ provides a simple example. For this family, the NEC, WEC,
and SEC are satisfied for $n\geq1$, while the DEC introduces the stronger
bound $n\geq3(\beta-2\pi)/(\beta-6\pi)$. The corresponding solutions are
regular at the throat, asymptotically flat, and traversable. The figures
confirm these analytical results and show clearly how the dominant energy
condition further restricts the allowed parameter space.

The distinction between the physical matter and the effective source is
central to the interpretation. The wormhole geometry still requires an
effective source that violates the null energy condition, as expected from
the flare-out condition. However, for $\beta>8\pi$ the physical and effective
null-energy combinations have opposite signs. The same branch also gives an
opposite-sign effective density for pressureless matter. Thus the trace
coupling shifts the exotic behaviour required by the geometry into the
effective gravitational sector rather than eliminating it. This makes the
solutions mathematically consistent but also shows that the weak-field
viability of the strong-coupling branch must be assessed separately. The
physical matter is likewise not separately conserved for $\beta\neq0$,
reflecting an exchange of energy--momentum with the gravitational sector.

A related global feature is that the complete admissible constant-redshift
class has zero ADM mass. Although the local physical energy density can be
positive, the bound imposed by the energy conditions forces $b(r)\to0$ at
infinity. This provides a useful asymptotic counterpart of the effective-source
interpretation above.

We also examined the marginal member $b(r)=r_0^2/r$, which saturates the
tangential NEC in the constant-redshift case. A decreasing redshift function
can make the tangential NEC strictly positive at the throat, but the global
analysis shows that this local improvement cannot be maintained throughout
an asymptotically flat exterior unless the redshift function is constant.
The explicit choice $\Phi=a r_0/r$ illustrates the same obstruction and
shows that sufficiently large redshift gradients can also drive the energy
density negative. Thus the additional freedom in $\Phi(r)$ does not rescue
the marginal $n=1$ configuration.

The present non-constant-redshift result is restricted to this boundary
member of the power-law family. It remains to determine whether solutions
with $n>1$ and a varying redshift function can preserve the energy conditions
globally. The dynamical stability of the non-exotic configurations is another
important open question. Because the theory contains a non-minimal
matter--torsion coupling, a consistent perturbative treatment should be
derived within $f(T,\tau)$ gravity rather than imported directly from general
relativity. These issues provide natural directions for extending the present
analysis.

\begin{acknowledgments}
This work was supported by the Deanship of Scientific Research, Vice
Presidency for Graduate Studies and Scientific Research, King Faisal
University, Saudi Arabia (Grant No.~KFU264529).
\end{acknowledgments}

\bibliographystyle{apsrev4-2}
\bibliography{WH_references}

\end{document}